\documentclass[reprint, amsmath,amssymb, aps, superscriptaddress, floatfix]{revtex4-1}

\usepackage[utf8]{inputenc}
\usepackage{graphicx}
\usepackage{dcolumn}
\usepackage{bm}
\usepackage{braket}
\usepackage{color}
\usepackage{units}
\usepackage{dsfont}
\usepackage{float}
\usepackage[colorlinks,linkcolor=blue,citecolor=blue,urlcolor=blue]{hyperref}
\usepackage{xcolor}

\usepackage{textcomp}

\begin{document}

\title{Full-field mode sorter using two optimized phase transformations \\
for high-dimensional quantum cryptography}

\author{Robert Fickler}
\affiliation{Institute for Quantum Optics and Quantum Information (IQOQI), Austrian Academy of Sciences, Boltzmanngasse 3, 1090 Vienna, Austria.}
\affiliation{Department of Physics, University of Ottawa, 25 Templeton Street, K1N 6N5, Ottawa, ON, Canada.}
\affiliation{Photonics Laboratory, Physics Unit, Tampere University, Tampere, FI-33720, Finland.}

\author{Fr\'ed\'eric Bouchard}
\affiliation{Department of Physics, University of Ottawa, 25 Templeton Street, K1N 6N5, Ottawa, ON, Canada.}

\author{Enno Giese}
\affiliation{Department of Physics, University of Ottawa, 25 Templeton Street, K1N 6N5, Ottawa, ON, Canada.}
\affiliation{Current address: Institut f\"ur Quantenphysik and Center for Integrated Quantum Science and Technology  (IQ\textsuperscript{ST}), Universit\"at Ulm, Albert-Einstein-Allee 11, 89081 Ulm, Germany}

\author{Vincenzo Grillo}
\affiliation{CNR-Istituto Nanoscienze, Centro S3, Via G Campi 213/a, I-41125 Modena, Italy.}

\author{Gerd Leuchs}
\affiliation{Department of Physics, University of Ottawa, 25 Templeton Street, K1N 6N5, Ottawa, ON, Canada.}
\affiliation{Max Planck Institute for the Science of Light, Staudtstr. 2, D-91058 Erlangen, Germany
and Institute of Optics, Information and Photonics, Department of Physics.}
\affiliation{Friedrich-Alexander-University Erlangen-Nuremberg, Staudtstr. 7/B2, D-91058 Erlangen, Germany.}

\author{Ebrahim Karimi}
\affiliation{Department of Physics, University of Ottawa, 25 Templeton Street, K1N 6N5, Ottawa, ON, Canada.}
\affiliation{Max Planck Institute for the Science of Light, Staudtstr. 2, D-91058 Erlangen, Germany
and Institute of Optics, Information and Photonics, Department of Physics.}

\begin{abstract}
High-dimensional encoding schemes have emerged as a novel way to perform quantum information tasks. For high dimensionality, temporal and transverse spatial modes of photons are the two paradigmatic degrees of freedom commonly used in such experiments. Nevertheless, general devices for multi-outcome measurements are still needed to take full advantage of the high-dimensional nature of encoding schemes. We propose a general full-field mode sorting scheme consisting only of up to two optimized phase elements based on evolutionary algorithms that allows for joint sorting of azimuthal and radial modes in a wide range of bases. We further study the performance of our scheme through simulations in the context of high-dimensional quantum cryptography, where high-fidelity measurement schemes are crucial.
\end{abstract}

\maketitle


\section*{Introduction}


Since its debut in 1984, quantum key distribution (QKD) has been one of the most considerable driving force of the broader field of quantum technologies~\cite{bennett1984quantum,gisin:02} and has been experimentally demonstrated in a wide range of optical configurations in the following decades~\cite{bennett:92,stucki2002quantum,schmitt2007experimental,liao:17}. Until recently, QKD has almost exclusively been realized with photonic \textit{qubits}, mainly due to their simple generation and detection. However, two-dimensional QKD systems, e.g. based on polarization or phase encoding, have their own limitations, such as the tolerable amount of noise in a channel or, equivalently, the distance of the link. In an attempt to overcome these limitations, \textit{high-dimensional} QKD was proposed using larger encoding alphabets~\cite{Bechmann-Pasquinucci:2000b,bourennane:2001}. High-dimensional QKD exploiting \textit{qudits} promises not only advantages in information capacity by encoding more than one bit of information per photon, but also in noise tolerance~\cite{Cerf2002,bouchard:17}. 

High-dimensional quantum information may be encoded using various photonic degrees of freedom. For instance, time bins, frequencies and transverse spatial modes are examples of high-dimensional encoding alphabets for photons. In particular, spatial modes of light have been recognized as a promising candidate for high-dimensional quantum information processing, due to their simplicity and versatility in generation, as well as their intrinsic phase stability. So far, most efforts have been directed towards a specific family of spatial modes consisting of beams carrying orbital angular momentum (OAM), also known as twisted photons~\cite{erhard2018twisted}. Demonstrations of high-dimensional QKD with twisted photons have been carried out in the laboratory~\cite{groblacher:06,mafu:13,mirhosseini:15,bouchard2018roundrobin,Bouchard2018experimental} and under realistic conditions~\cite{vallone2014free,Sit:17,bouchard2018underwater}. Moreover, they have also been an important tool in other quantum information tasks, such as quantum simulations or quantum entanglement verification~\cite{cardano:15,cardano:16,cardano:17,Bavaresco2018,erhard2018experimental}, to mention a few.
Hence, measuring spatial modes of light is crucial in high-dimensional quantum cryptography.

Single photons carrying OAM were first measured using a well-established technique known as \textit{phase-flattening}~\cite{mair2001entanglement}. Although possessing a mode-dependent bias~\cite{qassim:14}, it has become a standard tool in laboratories, requiring only a spatial light modulator (SLM) and a single mode fiber. However, this filtering technique consists of a projective measurement with an efficiency of $1/d$, where $d$ is the dimension. Therefore, in order to take full advantage of high-dimensional encoding schemes, a sorting-type of measurement, which allows $d$-outcome measurements, becomes necessary. Several of such OAM sorters have been proposed and realized in experiments using interferometric configurations~\cite{leach2002measuring} and diffractive elements~\cite{berkhout2010efficient,lavery:12,osullivan:12,mirhosseini2013efficient,ruffato:17,larocque2017generalized,walsh:18,saad:17}. Although efficient, these sorting schemes are inherently limited to sorting specific families of modes, in particular OAM beams and their discrete Fourier transform. 

To exploit the full potential of transverse spatial degree of freedom, one has to take the full-field mode structure of photons into account~\cite{salakhutdinov2012full,krenn2014generation} and, going beyond OAM, consider also radial modes~\cite{karimi2012radial,karimi2014radial,karimi2014exploring,plick2015physical}. Similar to phase-flattening, a technique called \textit{intensity-flattening} has recently been introduced to perform projective measurements on both azimuthal and radial modes of photons~\cite{bouchard2018measuring}. However, a full-field mode sorter is desirable for efficiently measuring spatial modes of photons in high-dimensional QKD. Other schemes based on scattering media~\cite{fickler2017custom} and interferometric configurations~\cite{zhou2017sorting,gu2018gouy} have been proposed and experimentally demonstrated to sort radial modes of light. Moreover, the simultaneous sorting of azimuthal and radial modes of light has recently been demonstrated using a technique employing multiple phase screens, namely \textit{wavefront matching}~\cite{fontaine2018optical,fontaine:17}. Although general, this technique requires a large number of independent phase elements. 

However, experimental constraints may not allow for the use of multiple phase screens, such that wavefront-matching methods cannot be applied. In this work, we investigate possible sorting mechanisms using only one or two phase modulations. Because there is in general no known procedure to design such phase elements, we employ an evolutionary optimization algorithm to find phase transformations that are custom-tailored to the modes to be sorted and the sorting geometry. Only when using two phase modulations, one in the near and one in the far field, we are able to sort both azimuthal and radial degrees of freedom of an incoming light field with nearly perfect distinction. Moreover, we demonstrate through our simulations that it should be also possible to use our scheme to sort all mutually unbiased basis, a task which has not been done before, such that the simulated phase patterns could be applied to QKD protocols. Although the presented scheme requires only two phase modulations, the near-perfect sorting comes at the cost of additional loss, which we found to scale inversely with the number of sorted modes. Hence, our result can be seen as a trade-off between experimental feasibility, sorting efficiency and acceptable levels of loss.

\begin{figure}
\centering
\includegraphics[width=0.48\textwidth]{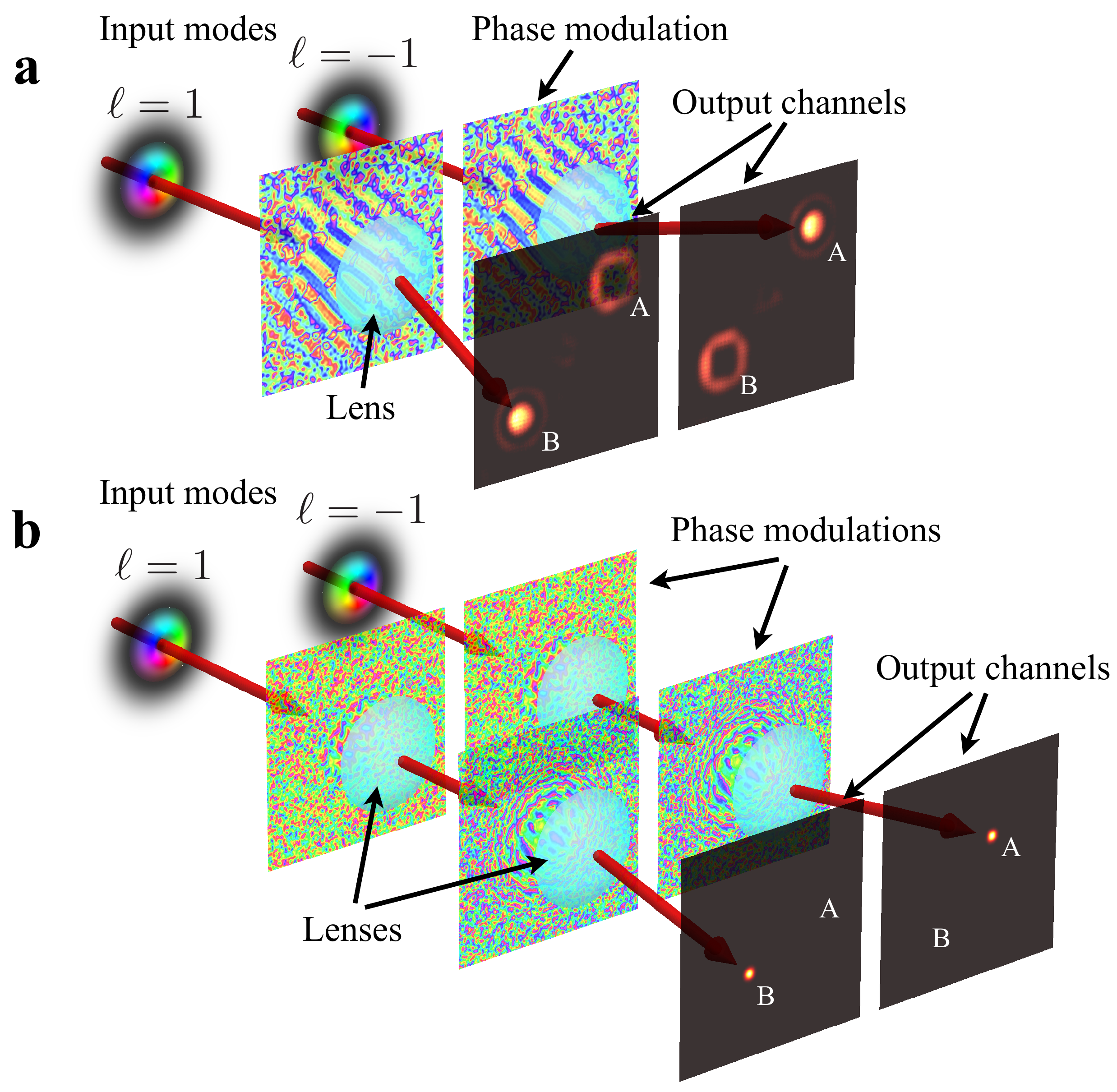}
\caption{Sketch of the simulated sorting using one \textbf{a} and two \textbf{b} optimized holograms. We optimize the holograms according to their ability to sort different input modes (here modes with +1 and -1 quanta of OAM) into a predefined output channel arrangement (here upper right and lower left corner).}
\label{fig:fig1}
\end{figure}

\begin{figure*}[!ht]
\centering
\includegraphics[width=1\textwidth]{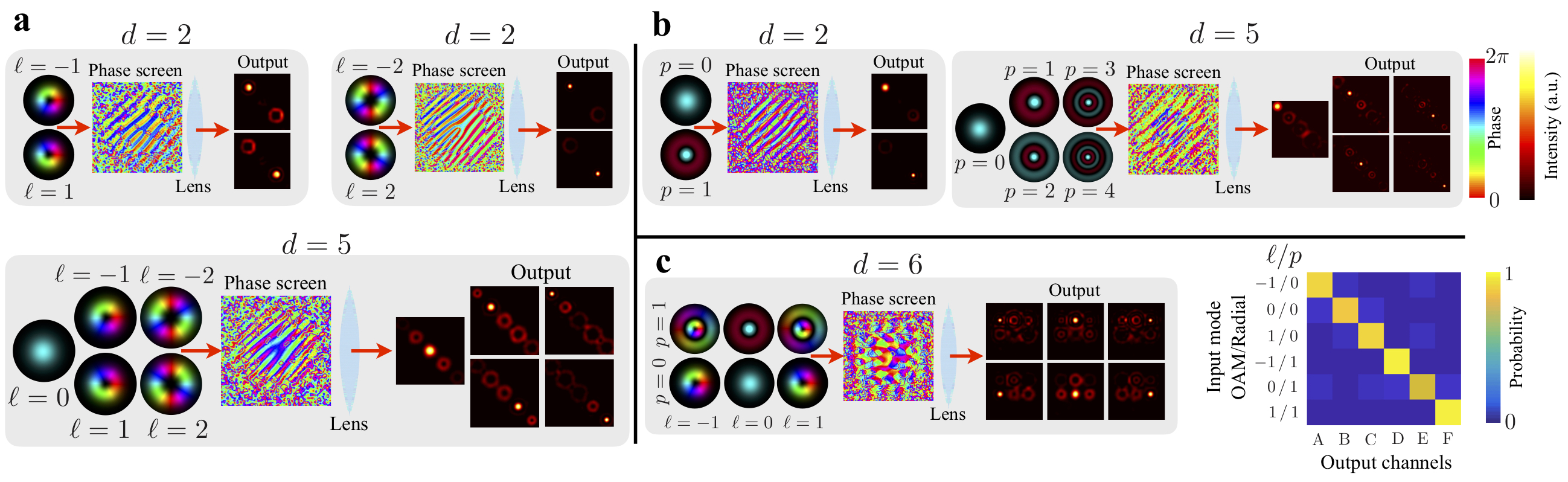}
\caption{Sorting of OAM, radial and full-field modes using a single optimized hologram. \textbf{a} The GA-designed holograms for OAM modes show superimposed grating structures with fork-like dislocations, thus, the found sorting technique is known from phase-flattening methods \cite{mair2001entanglement,gibson2004free,trichili2016optical}. \textbf{b} The holograms for radial modes similarly flatten the phase by taking the radial intensity profile into account. \textbf{c} The holograms for full-field modes, i.e. modes with higher order OAM and radial indices, show a more complex pattern, but still relies on a similar mechanism. We find in general that if more modes are included, shown for $d=5$ in \textbf{a} and \textbf{b} and for $d=6$ in \textbf{c}, the crosstalk between different output channels increases significantly. This can also be visually seen from an increase in incorrectly scattered light causing a reduction in efficiency and, to be more specififc, in the exemplary crosstalk matrix shown in \textbf{c} for full field sorting.}
\label{fig:fig2}
\end{figure*}

\section*{Evolutionary optimization procedure}

To develop a sorting mechanism that can be experimentally implemented in a straightforward manner, we design our optimization program to simulate laboratory conditions and limit the number of utilized phase elements to two (see Fig.~\ref{fig:fig1} for a sketch of the idea).
We perform all of our simulations with a wavelength of 780\,nm and a transverse spatial resolution of the phase elements of 20\,\textmu m, which is similar to the pixel size of commercially available SLMs, even though we could have also used other wavelength and phase elements. Our optimization procedure for one hologram sorting is as follows: each collimated mode is modulated by one phase element, an additional quadratic phase corresponding to a lens of 1\,m focal length is imprinted and a split-step method is used to propagate the beam to the focal plane. The overall sorting performance is evaluated by analyzing the intensities in the desired output channels, i.e. predefined spots of around 200\,\textmu m\,$\times$\,200\,\textmu m, for each mode. This ability to sort the freely chosen set of input modes is then optimized with the help of a genetic algorithm (GA). It first generates 10 different random phase patterns, each consisting of $125 \times 125$ macropixels in total, where (depending on the mode size) at least $40 \times 40$ pixels modulate the phase of the beam up to 2$\pi$. In addition, we blur the patterns with a Gaussian filter to prevent strong scattering that arises extreme phase gradients. These patterns are the population; and their fitness is evaluated according to their brightness in the $d$ desired output channels, a quantity we call sorting performance $B$. We define the latter as the intensity $I_n$ in the desired output channel $n$, from which we subtract the intensity in other $m$ output channels $\tilde{I}_m$: 
\begin{equation}
    B = \sum_{n=1}^d \left(I_n - \sum_{m\neq n} \tilde{I}_m\right).
\end{equation} 
 Hence, for each input mode $n$, we determine the intensity $I_n$ in the desired output channel as well as the intensities $\tilde{I}_m$ in the other incorrect $d-1$ output channels, from which we are then able to evaluate the sorting performance $B$.
 After this initialization phase that also contains a ranking of the phase patterns according to $B$, the so-called breeding is performed by combining two of the better phase elements of the population. They are chosen with a probability that exponentially decays according to their rank in the population. The two patterns are combined by randomly using one half of one and the complementary part of the other pattern. Additionally, up to 10\,\% of the macropixels (randomly-chosen) of the newly constructed pattern are mutated by about 15\,\%, before the fitness $B$ of the pattern is evaluated. This evaluation is performed by modulating all modes under consideration with the new phase pattern and by analyzing the sorting performance $B$ in the far field as described above.  If the last ranked phase pattern performs worse than the new pattern, it gets replaced. If the new pattern is not better, it is discarded. This procedure of breeding, analyzing and replacing is repeated with a slowly decreasing percentage of mutated macropixels (down to 0.01\,\%) until the sorting performance does not significantly improve. After the first $10^4$ iterations, we put an additional emphasis on low cross-talk by optimizing the fitness $F= B \cdot R$. In this definition, the sorting performance is multiplied with the secret-key rate 
\begin{equation}
    R = \log_2(d)-2h^{(d)}(e_b)
    \label{eq:keyrate}
\end{equation}
for high-dimensional states~\cite{sheridan2010security}, where $e_b$ is the quantum bit error rate (QBER), i.e. the normalized intensity found in the ``wrong'' output channels, and $h^{(d)}(x) := -x\log_2[x/(d-1)]-(1-x)\log_2(1-x)$ is the $d$-dimensional Shannon entropy. Using $F$ as the feedback signal during optimization, the algorithm not only maximizes the efficiency of the sorting but at the same time minimizes the cross-talk. The latter is especially important because we design holograms that are useful for quantum cryptography schemes, where only a very limited amount of cross-talk, i.e. errors, are permitted. The overall procedure remains the same in the sorting scheme using two holograms, however, each population member now consists of two phase modulations and the second hologram is placed in the focal plane of the first lens, such that we are able to also modulate the light field in momentum space.

\begin{figure*}
\centering
\includegraphics[width=0.95\textwidth]{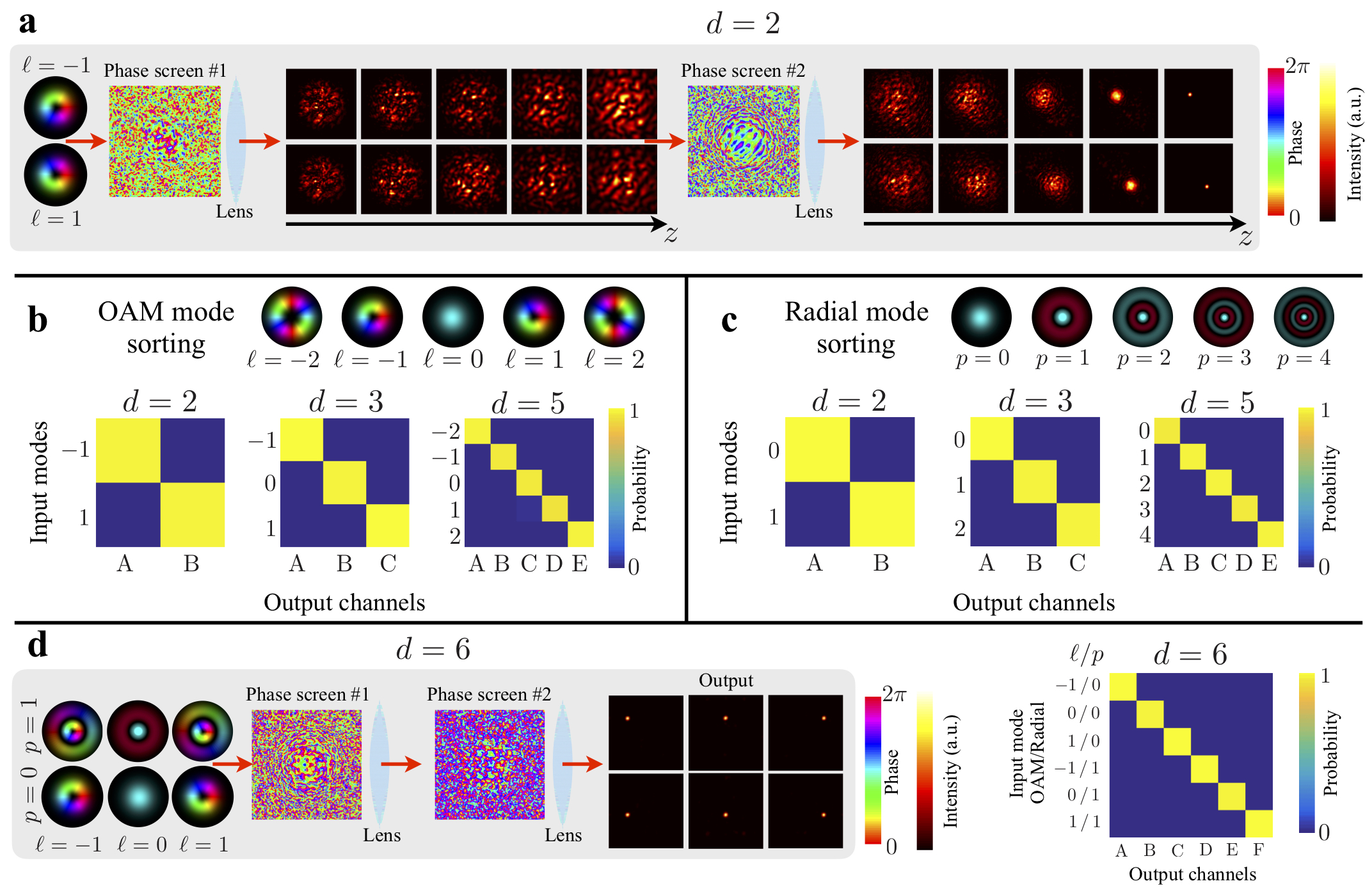}
\caption{Full-field sorting using two optimized holograms. \textbf{a} Propagation of LG modes with $p=0$ and $\ell=\pm1$ during the sorting process. We show the simulated intensity pattern for each mode in propagation steps of $\Delta z=$ 0.2 m. The first GA-designed phase modulation beaks the beam up and turns it into a mode-specific speckle pattern in the far field. The second phase modulation, placed in the far field, then imprints a phase such that the majority of the light is focused onto the predefined output-channels. Up to five OAM \textbf{b} and radial \textbf{c} modes can be sorted with near perfect distinction of $96-99$\,\% and an efficiency close to the $1/d$-limit, $d$ being the number of modes. \textbf{d} Full-field modes, i.e. modes with OAM and radial structure, can also be sorted with less than 1\,\% error.}
\label{fig:fig3}
\end{figure*}

\section*{Single hologram sorting}
At first we start by optimizing a single hologram and sorting Laguerre-Gauss (LG) modes of different OAM value $\ell$. We find that for up to five modes of different order (see Fig.~\ref{fig:fig2}-\textbf{a}) after approximately $10^5$ iterations, improvement of the sorting performance ceases. To quantify the performance of the resulting sorting we evaluate the normalized sorting probability $P_n$ for every mode $n$: $P_n = I_n / (I_n+\sum_{m\neq n}^d \tilde{I}_m)$ , where $I_n$ and $\tilde{I}_m$ stand for the intensities found in the predefined regions of the wanted and unwanted output channels, respectively. We refer to the mean value of the sorting probabilities of all modes as the sorting ability. For two modes, we find for two modes a sorting ability of around $98-99$\,\% with an efficiency of around $20-30$\,\%, irrespective of the OAM value (using $\ell = \pm 1$ and $\ell = \pm 2$). Interestingly, the obtained phase patterns shown in Fig.~\ref{fig:fig2}-\textbf{a} can be understood intuitively, in fact, they has already been used in various experiments~\cite{mair2001entanglement,gibson2004free,trichili2016optical} and are closely related to the standard technique of phase-flattening hologram (we have mentioned above): A grating diffracts the modes into different transverse regions and modulates the beam such that only one specific mode is transferred to a Gaussian spot in a specific diffraction order, i.e. the input mode is sorted. Starting from entirely random phase patterns, our optimization algorithm automatically finds the same sorting mechanism, where the diffraction orders of a grating correspond to the predefined output channels. We note that the GA-designed holograms show some minor imperfections such as speckle-like artifacts, which might lead to a slightly reduced efficiency but in principle could be removed by additional filtering techniques. Similar holographic patterns are also found when we optimize for five OAM modes ($\ell = 0, \pm 1, \pm 2$). The efficiency is still about 6\,\%, however, the sorting ability drops to $90 \pm 4$\,\%. 

As a second step, we sort modes of different radial index $p$, see Fig.~\ref{fig:fig2}-\textbf{b}. We start by optimizing the sorting performance $B$ of two different modes ($p= 0,1$) and find the same mechanism, i.e. structured gratings that lead to Gaussian-like spots in specific diffraction orders corresponding to the predefined sorting channels. Again, the cross-talk between the modes is low with a sorting ability $P_n$ of $98-99$\,\%. We continue with sorting the five lowest radial modes ($p=0-4$) and again achieve values similar to the ones found for OAM modes, i.e. a sorting ability of $89 \pm 1$\,\% with an average efficiency of 7\,\%. As a final task using a single hologram we sort $d=6$ modes that are a combination of both characteristic indices $\ell=0,\pm 1$ and $p=0,1$, see Fig.~\ref{fig:fig2}-\textbf{c}. 
Although the average sorting ability is $90 \pm 6$\,\% and remains roughly the same, we find a stronger variation between different sorting probabilities and a further reduction of the efficiency efficiency to less than 5\,\%. This decrease in the quality of sorting hints towards the limitations of using a single phase element. Nevertheless, this set of simulations underlines that the optimization procedure converges to complex solutions, i.e. it finds well-established sorting mechanisms corresponding to known optical elements starting from a random phase pattern.

\section*{Sorting using two holograms}

As a natural extension we investigate the performance of our approach when additionally a second phase modulation occurs in the far field. Similar to the previous section, the modes are modulated by one phase element, put into the initially collimated beam. With the help of a quadratic phase term simulating a lens with a focal length of 1\,m, the light is then approximately brought to the Fourier plane by propagating it over this distance with a split-step method. In this plane we introduce a second phase modulation of the light field. Again, we additionally imprint a quadratic phase term with focal length of 1\,m, propagate the field for this distance, and investigate the resulting intensity pattern for each mode one by one. Analog to the feedback signal described above, we define well-separated output channels (200\,\textmu m\,$\times$\,200\,\textmu m) for each input mode, and use the sorting performance $B$ as a feedback signal for the optimization. Our GA modulates the phase both in the near and in the far field, starting from random phases. As a first set of simulations we investigate the sorting ability of up to five LG modes, which differ only in their OAM values (see Fig.~\ref{fig:fig3}-\textbf{a}-\textbf{b}). For two ($\ell = \pm 1$), three ($\ell = 0, \pm 1$) and five modes ($\ell = 0,\pm 1, \pm 2$) we find after approximately $10^5$ iterations a sorting ability of $99.6 \pm 0.1$\,\%, $99.3 \pm 0.3$\,\%, and $96.8 \pm 1.5$\,\% with efficiencies of around 40\,\%, 25\,\%, and 15\,\%, respectively. Interestingly, all results seem to approach a $1/d$-limit, where $d$ stands for the number of modes. In contrast to a single phase modulation, there is no intuitive explanation of the underlying sorting mechanism based on gratings. The first phase modulation appears to break up the beam to generate a complex mode-dependent speckle pattern. The second phase modulation then leads to a refocusing of the beam at the predefined spot (see Fig.~\ref{fig:fig3}-\textbf{a} for an exemplary propagation during the sorting of $\ell=\pm 1$), while the rest of the light is scattered to such a broad region that it is not visible anymore. Thus, compared to the adiabatic approach towards sorting described in~\cite{fontaine:17,fontaine2018optical}, our method relies only on two holograms and achieves minimal cross-talk at the cost of high loss due to the strong scattering caused by phase modulations of high spatial frequencies.  Strong scattering implies a high sensitivity to alignment so that the experimental implementation has to be done with great care and accuracy.
At the same time, it is also the main reason for the enormous flexibility of the scheme, since we do not find any dependence of performance on the specific modal set or the output channel geometry under consideration.

In a second set of simulations, we further verify the possibility to custom-tailor the sorting scheme by optimizing for up to five radial modes without any reduction of its sorting ability, see Fig.~\ref{fig:fig3}-\textbf{c}. For two ($p=0,1$), three ($p=0-2$), and five ($p=0-4$) different modes, we find sorting abilities of $99.5 \pm 0.4$\,\%, $99.8 \pm 0.1$\,\%, and $97.7 \pm 0.7$\,\%, respectively, with effiencies up to 40\,\%, 25\,\% and 15\,\%, thus, again approaching the  $1/d$-limit. 

After studying azimuthal as well as radial modes separately, we test the full capability of our scheme by sorting any combination of transverse spatial modes. We optimize for the sorting of six modes using a both degrees of freedom, where the OAM degree of freedom ($\ell=0, \pm 1$) is sorted horizontally and the radial degree of freedom ($p=0,1$) vertically. With this geometry, we take advantage of the full two-dimensional state-space of transverse modes. After roughly $2\times 10^5$ iterations, we find a sorting ability of $99.3\pm0.1$\,\% with an efficiency of roughly 15\,\% (see Fig.~\ref{fig:fig3}-\textbf{d}). Thus, our approach is able to perfectly decompose any light field into its full-field components, limited only by the observed $1/d$-efficiency.

\begin{figure}
\centering
\includegraphics[width=0.48\textwidth]{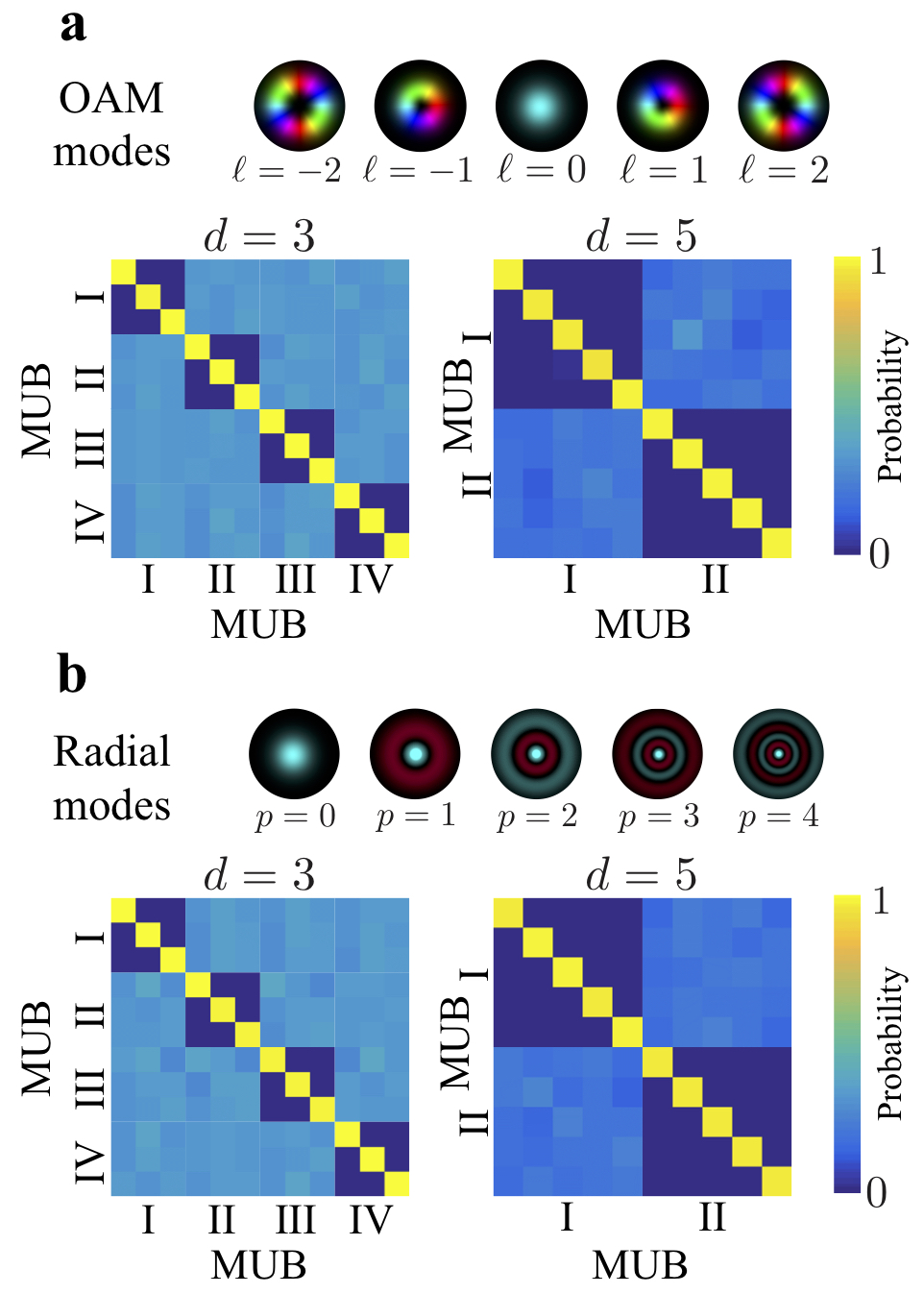}
\caption{Sorting of mutually unbiased bases (MUB). \textbf{a} Sorting of all four MUBs for three-dimensional states (left) and two MUBs for five-dimensional states encoded in OAM (right). \textbf{b} Analogous sorting for radial modes, i.e. sorting of all four MUBs for three-dimensional states (left) and two MUBs for five-dimensional states (right). }
\label{fig:fig4}
\end{figure}

\section*{Application to high-dimensional quantum cryptography}

As a final demonstration of the broad applicability of the proposed scheme, we study the sorting of complex superpositions of the set of modes used above, i.e. their mutually unbiased basis (MUB). These bases are crucial for quantum communication as they are required for QKD schemes, where the generation basis at the sender as well as the measurement basis at the receiver is randomly switched between at least two of them. By simply replacing the LG modes by their MUBs, it is straightforward to implement their sorting. Not surprisingly, we find very similar results for all possible bases we studied. For the three additional MUBs of the modes of $\ell=0,\pm 1$, we find $99.6\pm0.2$\,\%, $99.6\pm0.1$\,\%, and $99.8\pm0.1$\,\% for their sorting abilities. Using Eq.~\eqref{eq:keyrate}, this very low QBER of $e_b =0.4$\,\% translates into an achievable secure key rate of 1.50\,bits per sifted and detected photon.
For radial modes of $p=0,1,2$ the results are very similar. The sorting abilities $98.9\pm0.4$\,\%, $98.7\pm0.2$\,\%, and $99.5\pm0.1$\,\% lead to an overall QBER of $e_b=0.8$\,\% and a key rate of 1.44\,bits per sifted photon. 
Importantly, we also find a near-perfect unbiasedness when sending states from one basis through the phase modulations optimized for another MUB (see cross-talk matrices in Fig.~\ref{fig:fig4} on the left), i.e. all channels show a detection efficiency of $33\pm1$\,\% for OAM and $33\pm2$\,\% for radial modes, respectively.

Finally, we simulate sorting of a second MUB for five-dimensional quantum states encoded in OAM ($\ell=0,\pm 1 , \pm 2$) and radial modes ($p=0 -4$). As before, we find almost the same sorting ability of $98.5\pm0.2$\,\% and $97.2\pm0.3$\,\% for the second MUB for OAM and radial modes, respectively. Because the error remains very low, i.e. $e_b=2.32$\,\% and 2.55\,\%, the possible key rate for this five-dimensional QKD scheme would be 1.91 (OAM modes) and 1.87 (radial modes) bits per sifted and detected photon. The cross talk matrices are shown on the right of Fig~\ref{fig:fig4}. In general, our approach allows for the implementation of passively random switching between all utilized MUBs by simply adding appropriate beam splitters followed by the two optimized holograms and appropriate detector arrangements (similar to~\cite{mirhosseini:15}). 

\section*{Conclusions and outlook}
In conclusion, we have demonstrated method to realize a sorter for transverse spatial modes using only two phase modulations. At the cost of losses, it is possible to sort up to at least six modes of lower orders of the full optical field, i.e. modes characterized by azimuthal and radial degrees of freedom. Furthermore, our scheme can be applied to quantum experiments, in particular quantum cryptography, where sorting of different MUBs with very low cross-talk is necessary. Because our method allows for a flexible positioning and adjusting of output channels and only requires two phase modulations, it can be straightforwardly applied to a broad range of experimental situations. It is possible to either use offline-designed static phase modulations or to optimize actively in an experiment with computer-controlled SLMs and the output of appropriate detectors as the feedback signal~\cite{fickler2017custom}. The latter scheme also automatically compensates for experimental imperfections and as such only requires a stable setup. On the other hand, using offline-designed phase modulations can be realized by refractive optical elements thereby minimizing additional losses with the additional challenges of requiring a high precision of alignment along with a aberration-free optical beam path. Finally, our scheme could be extended to sort non-orthogonal states~\cite{bent2015experimental}, for state tomography and discrimination schemes.

\section*{Acknowledgements}
R.F acknowledges support from the Austrian Science Fund (FWF) through the START project Y879-N27, the Banting Postdoctoral fellowship of the NSERC and the Academy of Finland (Competitive Funding to Strengthen University Research Profiles - decision 301820 and Photonics Research and Innovation Flagship - decision 320165). F.B. acknowledges the financial support of the Vanier graduate scholarship of the NSERC. This work was supported by the Ontario's Early Researcher Award (ERA), European Union's Horizon 2020 Research and Innovation Programme (Q- SORT) grant number 766970, Canada Research Chairs (CRC) and Natural Sciences and Engineering Research Council of Canada (NSERC).

\bibliography{apssamp}

\end{document}